# Role of 1-D finite size Heisenberg chain in increasing metal to insulator transition temperature in hole rich $VO_2$

Raktima Basu,[1,*] Manas Sardar,[2] Santanu Bera,[3] P. Magudapathy,[2] Sandip Dhara[1,*]

[1]Nanomaterials and Sensor Section, Surface and Nanoscience Division, Indira Gandhi Centre for Atomic Research, Homi Bhabha National Institute, Kalpakkam-603102, India

[2]Materials Physics Division, Indira Gandhi Centre for Atomic Research, Kalpakkam-603102, India

[3]Water and Steam Chemistry Division, BARC Facility, Kalpakkam-603102 & Homi Bhabha National Institute, Mumbai-400094, India

E-mail: raktimabasu14@gmail.com ; dhara@igcar.gov.in

Abstract

$VO_2$ samples are grown with different oxygen concentrations leading to different monoclinic, M1 and triclinic, T insulating phases which undergo a first order metal to insulator transition (MIT) followed by a structural phase transition (SPT) to rutile tetragonal phase. The metal insulator transition temperature ($T_c$) was found to be increased with increasing native defects. Vanadium vacancy ($V_V$) is envisaged to create local strains in the lattice which prevents twisting of the V-V dimers promoting metastable monoclinic, M2 and T phases at intermediate temperatures. It is argued that MIT is driven by strong electronic correlation. The low temperature insulating phase can be considered as a collection of one-dimensional (1-D) half-filled band, which undergoes Mott transition to 1-D infinitely long Heisenberg spin ½ chains leading to structural distortion due to spin-phonon coupling. Presence of $V_V$ creates localized holes ($d^0$) in the nearest neighbor, thereby fragmenting the spin ½ chains at nanoscale, which in turn increase the $T_c$ value more than that of an infinitely long one. The $T_c$ value scales inversely with the average size of fragmented Heisenberg spin ½ chains following a critical exponent of ⅔, which is exactly the same predicted theoretically for Heisenberg spin ½ chain at nanoscale undergoing SPT (spin-Peierls transition). Thus, the observation of MIT and SPT at the same time in $VO_2$ can be explained from our phenomenological model of reduced 1-D Heisenberg spin ½ chains. The reported increase (decrease) in $T_c$ value of $VO_2$ by doping with metal having valency less (more) than four, can also be understood easily with our unified model, for the first time, considering finite size scaling of Heisenberg chains.

# Introduction

Among all oxides of vanadium, $VO_2$ finds tremendous attraction and turns out to be the center of extensive research because of its first order, reversible metal to insulator transition (MIT) at the technologically important transition temperature of 340K, which is very close to room temperature.[1] A structural phase transition (SPT) from a low temperature monoclinic M1 (space group $P2_1/c$) to a high temperature rutile tetragonal R (space group $P4_2/mnm$) is also associated with the MIT.[2,3] Besides M1, another two low temperature phases of monoclinic M2 (space group $C2/m$) and triclinic T (or monoclinic M3; space group $\overline{C}1$) are also reported to evolve during the phase transition from M1 to R.[4] The M2 and T phases can be stabilized at room temperature by doping with metals of lower valency than $V^{4+}$.[5-7] The phase stabilization of the metastable phases is alternatively reported by applying tensile strain along the rutile $c$ axis ($c_R$).[8,9] In each phase of $VO_2$ there are two interpenetrating parallel chains of V surrounded by six O atoms forming distorted octahedron. In the high temperature R phase, all the V chains are parallel and periodic,[10] whereas in the low temperature monoclinic M1, M2 and triclinic T phases there are significant differences in the arrangement of V along $c_R$ axis. The V forms pair (dimerized) and the pairs tilt along the $c_R$ axis in the M1 phase, making the unit cell double of the unit cell in R phase (schematic in supplementary information Fig. S1) with the approximate crystallographic relationship as $a_{M1} \leftrightarrow 2c_R$, $b_{M1} \leftrightarrow a_R$, and $c_{M1} \leftrightarrow b_R - c_R$.[5,11] In M2 phase, the one set of V chains along the $c_R$ axis pair without twisting, while the V ions in the nearest neighbor V chains, do not pair but twist away from $c_R$ axis. The crystallographic relationship with the R phase is $b_{M2} \leftrightarrow 2c_R$, $a_{M2} \leftrightarrow 2a_R$, and $c_{M2} \leftrightarrow -b_R$.[11] The triclinic T phase is intermediate between the M1 and M2 phases, in which paired V chains in the M2 phase are reported to be twisted slightly.[12] In addition to temperature, other driving forces such as external electric field,[13] hydrostatic pressure,[14] intense illumination,[15] and strain[16] can also introduce phase transition in $VO_2$. Moreover, the transition temperature ($T_c$) is also reported to be tuned by altering carrier density,[17] applying strain,[18] or doping[19] with significant change in the optical, thermal and electrical properties of $VO_2$. These characteristics make $VO_2$ a promising material for various electrical, thermal and optical devices such as smart windows,[20] gas sensors,[21] electrical switches,[22] and cathodes for Li-ion batteries.[23] Doping in bulk $M_xV_{(1-x)}O_2$ can lead to decrease[24] in the $T_c$ value for M = $W^{+6}$, $Mo^{+6}$, $Ta^{+5}$, $Nb^{+5}$ and an increase[5-7] is reported in the $T_c$ value for M = $Al^{+3}$, $Ga^{+3}$, $Cr^{+3}$.

Since the MIT is generally accompanied by a SPT in these materials, the controversy remains, whether the residual lattice strain and dynamical electron (spin)-phonon coupling or strong electron-electron correlation triggers the MIT.[25,26] We argue that random local strains due to vanadium vacancies ($V_V$) leads to evolution of M1 phase into M2 or T phase or a mixture of them, mainly because the random local strains prevents cooperative twisting of the V-V dimers in the spin-Peierls state. However, they do not force the MIT.

In the present study, we address the origin of MIT as well as the effect of defects in changing the transition temperature of $VO_2$ samples invoking Coulomb correlation and dimensional reduction in V chains. The existence of strain, introduced in presence of the native defect because of the local deviation in the $VO_2$ stoichiometry is confirmed by Raman spectroscopic analysis and glancing angle X-ray diffraction (GIXRD) measurements. The dimensional reduction leads to Mott-Hubbard MIT and the insulating phase of $VO_2$ can be considered as one-dimensional (1-D) non-interacting Heisenberg spin ½ chains undergoing Peierls type SPT. The role of $V_V$ in increasing the transition temperature is discussed in details using X-ray photoelectron spectroscopic (XPS) studies and introducing finite size 1-D Heisenberg spin ½ chain model in the hole rich (acceptor doped) system.

**Experimental details**

$VO_2$ nanostructures were grown on crystalline (*c*-)Si(111) substrate by vapor transport process using bulk $VO_2$ powder (Sigma-Aldrich, 99%) as source and Ar (99.9%) as the carrier gas. The bulk $VO_2$ powder was placed in a high pure (99.99%) alumina boat inside a quartz tube. The reaction chamber was kept in a furnace and was pre-evacuated up to $10^{-3}$ mbar. Substrate was kept 10 mm away from the source and normal to the stream of carrier gas. The temperature of the quartz tube was programmed to rise up to the optimized growth temperature (1150K) with a ramp rate of 15K min$^{-1}$. The synthesis was carried for 4h with different optimized O exposure by flowing 10, 20, 30 and 40 sccm (sample, $S_1$, $S_2$, $S_3$ and $S_4$, respectively) of Ar (99.9%) containing ≈$2\times10^5$ ppm O.

The crystallographic orientation and phase confirmation studies were carried out with the help of GIXRD (Inel, Eqinox 2000) using a Cu Kα radiation source of wavelength, λ=1.5406 Å with a glancing angle (θ) of 0.5° in the θ-2θ mode. The XPS (VG ESCALAB MK200X) analysis was performed for the $VO_2$ samples synthesized at different growth conditions using an X-ray source of Al-Kα (1486.6 eV) with beam diameter around 3 mm and collection area (with the largest slit) approximately 2x3 mm$^2$. The binding energy (BE) values were measured with respect to the C 1*s* reference peak. The spectra were processed by applying Shirley type background and the curves were fitted by mixture of Gaussian–Lorentzian line shapes. The vibrational modes of the synthesized samples were analyzed using a micro-Raman spectrometer (inVia, Renishaw, UK) in the backscattering configuration with Ar$^+$ Laser (514.5 nm) as the excitation source, diffraction gratings of 1800 gr.mm$^{-1}$ as monochromator and a thermoelectrically cooled charged couple device (CCD) as the detector.

## Results and discussions

*Crystallographic structural studies*

The X-ray crystallographic structural studies of the samples with minimum ($S_1$) and maximum ($S_4$) exposure to O are shown in Figure 1. In sample $S_1$, the diffraction peak at 2θ = 27.83° corresponds to the (011) plane of monoclinic M1 phase (equivalent to (110) plane of R phase)[27] of $VO_2$ (JCPDS # 04-007-1466), which is the preferential growth plane for $VO_2$.

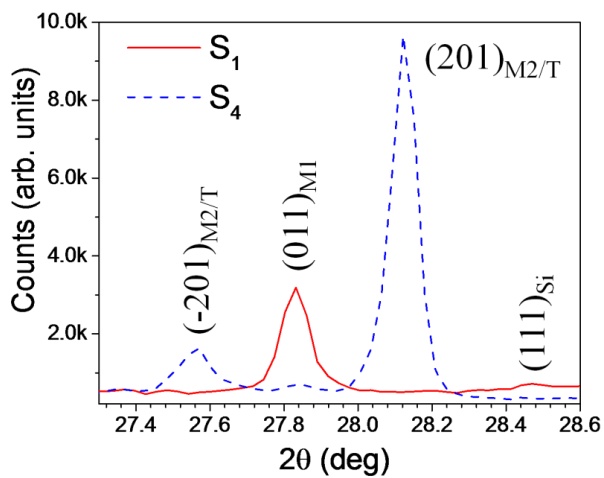

**Fig. 1.** GIXRD spectra of the pristine samples (a) $S_1$ and (b) $S_4$ indicating crystallographic (*hkl*) planes of the corresponding phases.

In sample $S_4$, two peaks are observed at 2θ = 27.56° and 28.12°, which may correspond to (-201) and (201) planes of either M2 phase (JCPDS # 00-033-1441) or T (M3) phase (JCPDS # 01-071-0289) of $VO_2$.[28,29] The presence of a less intense peak at 2θ = 27.84° in sample $S_4$ denotes coexistence of M1 phase with a (011) lattice plane compression. Compression in one direction generally is coupled to an expansion in the perpendicular direction. Thus, the compression in $(011)_{M1}$ plane leads to tensile strain along $c_R$ axis, and can stabilize M2 or T phase.[9] In the present study, the role of substrate[30] in introducing strain in these samples is ruled out as they are synthesized on the same substrate keeping all growth parameters identical, except for the amount of gas exposure. Since, the samples were grown for different percentage of flow of Ar (containing ≈$2\times10^5$ ppm O); there would be departure from perfect stoichiometry in $VO_2$, because of different amounts of O present in the carrier gas. Presence of excess O in the sample $S_4$ is probable, as it is grown with highest amount of O exposure at high temperature (1150K). The excess O creates $V_V$ for maintaining the charge neutrality, by the defect reaction (in Kroger-Vink notation), $O_2$ => $V_V^{4+} + 4h^+ + 2O^{-2}$, where $O^{-2}$ are oxygen anions in the lattice and the four holes, $h^+$ are created at the V sites adjacent to the $V_v$ site. These holes are trapped at the V sites of the neighboring chains by converting $V^{4+}$ to $V^{5+}$ or $d^0$ (spin $S=0$) states leading to acceptor doping.[31] $V^{4+}$ is located at the center of the oxygen octahedron with principal axes perpendicular to $(110)_{M1}$ lattice plane.[12] Thus, V–O bond length is reduced as the $V^{4+}$ is replaced by $V^{5+}$, and two apical $O^{2-}$ of the octahedron shift closer to each other, which in turn reduce the (110) lattice plane spacing of M1 phase. The shortening of V-O bonds around each $V^{5+}$ ion also reduces the twisting of $V^{4+}$-$V^{4+}$ dimers (the spin-Peirls singlets of the $S=½$ chains) away from the $c_R$ axis in its neighborhoods, and can produce M2 or T phase. The presence of M2 or T phase and trace of M1 phase in the O rich sample ($S_4$) suggests this picture (Fig. 1).

*Spectroscopic studies for phase transition*

Raman spectroscopic analysis was carried out to acquire further information of the phases present in the pristine samples. The typical Raman spectra for all the four samples $S_1$ to $S_4$ at room temperature are shown in Figure 2(a). Group theoretical analysis predicts eighteen Raman-active phonon modes for all low temperature $VO_2$ phases of M1: $9A_g+9B_g$, and for M2 and T: $10A_g+8B_g$ at $\Gamma$ point but with different symmetries.[14] However, we report twelve vibrational modes for sample $S_1$ (Fig. 2(a)).

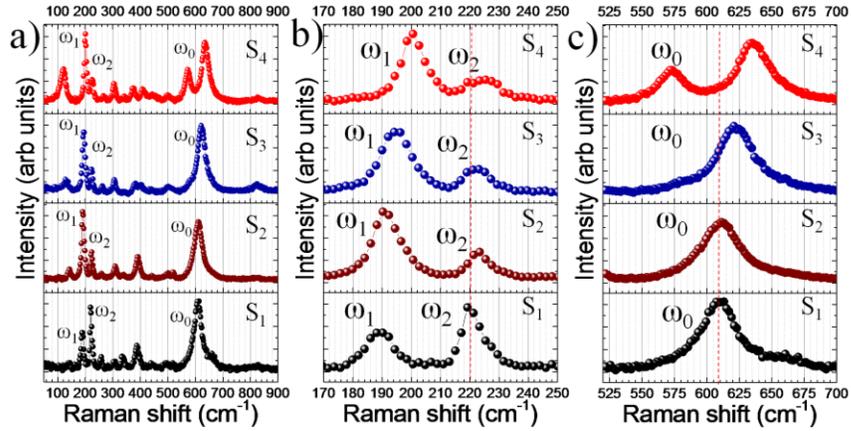

**Fig. 2.** (a) Raman spectra of the pristine samples, $S_1$ to $S_4$ with proper symmetry notations, (b) Raman modes correspond to V-V chain ($\omega_1$ and $\omega_2$) for all the four samples and (c) characteristic peak of $VO_2$ ($\omega_0$) corresponds to V-O stretching for the samples.

Raman modes at 141 ($A_g$), 189 ($A_g$), 220 ($A_g$), 258 (either $A_g$ or $B_g$; $A_g/B_g$), 310 ($A_g$), 336 ($A_g$), 388 ($A_g/B_g$), 440 ($A_g/B_g$), 494($A_g/B_g$), 609($A_g$), 665($B_g$), 826($B_g$) cm$^{-1}$ confirm the presence of pure M1 phase of $VO_2$ in sample $S_1$ and is in good agreement with the previously reported data.[32,33] All Raman modes in sample $S_1$ except 141, 440 and 826 cm$^{-1}$ are observed to be blue shifted by an amount of ~2-3 cm$^{-1}$ in sample $S_2$. In case of sample $S_3$, we report thirteen Raman modes at 130 ($A_g$), 195 ($A_g$), 223 ($A_g$), 262 ($A_g/B_g$), 306 ($A_g$), 340 ($A_g$), 381 ($A_g/B_g$), 403 ($A_g/B_g$), 440 ($A_g/B_g$), 501 ($A_g/B_g$), 578, 622 ($A_g$), and 828 ($B_g$) cm$^{-1}$. The thirteen Raman active modes in sample $S_4$ are observed at 121 ($A_g$), 201 ($A_g$), 225 ($A_g$), 267 ($A_g/B_g$), 304 ($A_g$), 343 ($A_g$), 374 ($A_g/B_g$), 409 ($A_g/B_g$), 440 ($A_g/B_g$), 501($A_g/B_g$), 572, 636 ($A_g$), and 828 ($B_g$) cm$^{-1}$, which mostly resemble with reported data for T phase of $VO_2$.[4,17] The two major differences are observed in the Raman spectrum of the sample $S_4$ as compared to the sample $S_1$; (i) significant blue shift of the Raman modes is observed at 189 cm$^{-1}$ ($\omega_1$), and 220 cm$^{-1}$ ($\omega_2$) by an amount of 12 and 5 cm$^{-1}$, respectively (Fig. 2(b)) and (ii) splitting of the Raman mode is observed for 609 cm$^{-1}$ ($\omega_0$) into two peaks at 572 and 636 cm$^{-1}$ (Fig. 2(c)), which is a signature of presence of the T phase.[14] The Raman modes of $\omega_1$ and $\omega_2$ correspond to the vibration of V ions in the transverse and along the *c* axis, respectively. Whereas, Raman mode of $\omega_0$ arises due to V-O vibrations.[14] Strain induced blue shift (phonon hardening) of $\omega_0$, $\omega_1$ and $\omega_2$ modes were reported earlier.[4] In our samples ($S_2$, $S_3$ and $S_4$), the interfacial strain with substrate does not vary. It implicates that the hardening of these modes may also happen due to presence of local strains around $V_v$, as well as shortening of $V^{+5}$-$O^{-2}$ distances, even when they are present in small concentrations. The excess O in samples $S_2$, $S_3$ and $S_4$ has two effects, (i) reduction in the amount of twisting of dimers and stabilizing M2/T phases, and (ii) modifying the V-V chain distance resulting in blue shift of the $\omega_0$, $\omega_1$ and $\omega_2$ modes.

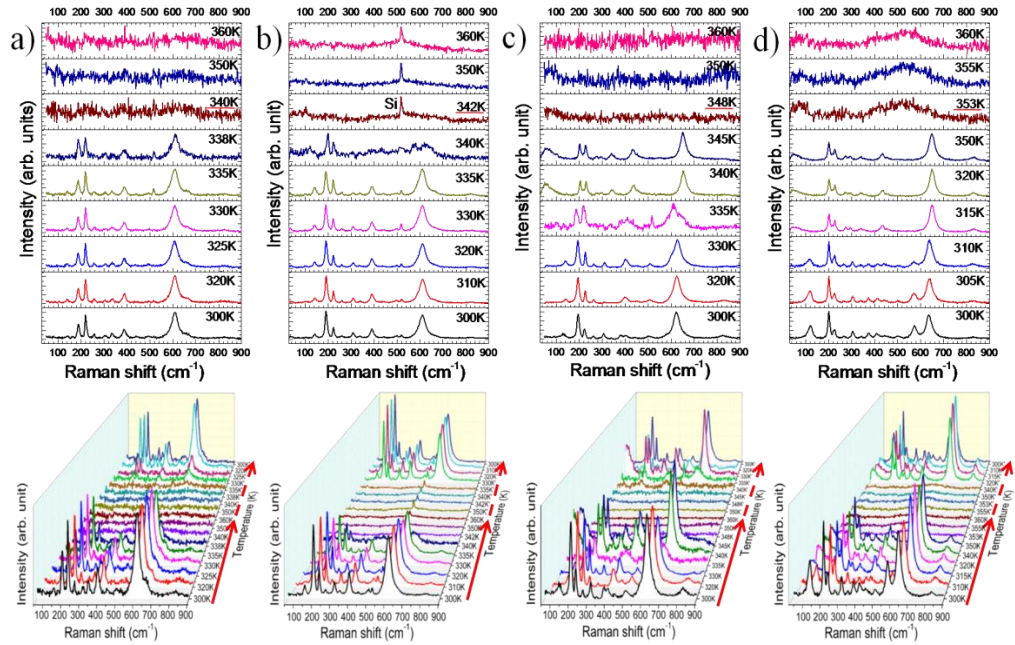

**Fig. 3.** Raman spectra of the samples (a) $S_1$, (b) $S_2$ (c) $S_3$ and (d) $S_4$ as a function of temperature. Underlined temperatures denote the corresponding transition temperatures and phase reversibility. Peak observed at 521 cm$^{-1}$ in (b) for sample $S_2$ above 340K correspond to Si substrate. Solid and dashed arrows denote the increase and decrease in temperature, respectively.

Figure 3 shows the Raman spectra of the samples, $S_1$ to $S_4$ in the temperature range from 300 to 360K using Linkam (THMS600) stage. All the Raman modes disappear at 340, 342, 348 and 353K in sample $S_1$ (Fig. 3(a)), $S_2$ (Fig. 3(b)), $S_3$ (Fig. 3(c)), and $S_4$ (Fig. 3(d)), respectively, confirming the transition to metallic R phase for these samples at different temperatures. Going from $S_1$ to $S_4$, the $T_c$ value increases by 13 K. In sample $S_3$, a noticeable change is observed above 335K. The Raman peak observed at 622 cm$^{-1}$ ($\omega_0$) splits into two peaks at 611 and 646 cm$^{-1}$, confirming the coexistence of $M_1$ and $M_2$ phase above 335

K.[4,14] At 340K, the Raman peaks indicate pure M2 phase with strong V-O peak ($\omega_0$) at 650 cm$^{-1}$ and V-V peaks at 204 ($\omega_1$) and 229 cm$^{-1}$ ($\omega_2$).[4,14] All Raman modes, however disappeared at 348K leading to a transition from M2→R phase. So for S$_3$, the phase evolution is T→M2→R with the rise in temperature. In sample S$_4$, the T→M2 phase transition is observed at 320K and M2→R at 353K.

In order to understand the phenomenological origin of the increase in $T_c$ with increase in oxygen exposure, XPS studies were performed for all the samples. The Shirley type background corrected XPS spectra for different elements and their characteristic electronic transitions for sample S$_1$ to S$_4$ are shown in Figure 4.

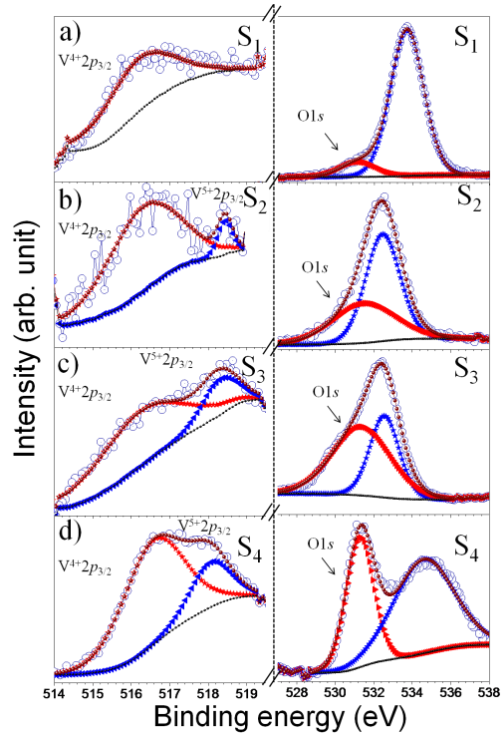

**Fig. 4.** High resolution XPS spectra of sample (a) S$_1$, (b) S$_2$, (c) S$_3$, and (d) S$_4$ with corresponding electronic transition of different elements. Open circles represent the data points, solid arrows and circles represent the fitted curves. In the left panel curves with red and blue arrows correspond to V$^{4+}$ and V$^{5+}$ oxidization states, respectively. In the right panel red arrows correspond to lattice oxygen, and blue arrows corresponds to oxygen from SiO$_2$ or absorbed oxygen species.

For sample S$_1$, V 2$p_{3/2}$ spin-orbit spectrum (Fig. 4(a)) can be fitted into single peak with the BE value 516.2 eV and is assigned to V$^{4+}$ oxidization state.[34] In sample S$_2$, the spin-orbit spectrum can be fitted by two peaks with BE values 516.3 and 518.4 eV (Fig. 4(b)). The V 2$p_{3/2}$ peak, observed at lower BE value of 516.3 eV for sample S$_2$, can be assigned to V$^{4+}$ oxidization state. The peak identified at higher BE value of 518.4 eV can be assigned to V$^{5+}$ oxidization state.[35] These peaks are observed at 516.3 eV (V$^{4+}$) and 518.3 eV ($V^{5+}$) in sample $S_3$ (Fig. 4(c)) and at 516.3 eV ($V^{4+}$) and 518.2 eV ($V^{5+}$) in sample $S_4$ (Fig. 4(d)). The ratio of area under the curves of the two peaks can be a quantitative measure of the ratio $V^{5+}/V^{4+}$. This ratio increases progressively from sample $S_1$ to sample $S_4$. Table I contains the positions of the corresponding peaks and $V^{5+}/V^{4+}$ ratio for all the four samples.

**Table I:** Binding energy values of $V^{4+}$ and $V^{5+}$ and $V^{5+}/V^{4+}$ ratio for samples $S_1$ to $S_4$

| Sample | $V^{4+}$ (in eV) | $V^{5+}$ (in eV) | $V^{5+}/V^{4+}$ |
|---|---|---|---|
| $S_1$ | 516.2 | - | 0 |
| $S_2$ | 516.3 | 518.4 | 0.083 |
| $S_3$ | 516.3 | 518.3 | 0.215 |
| $S_4$ | 516.3 | 518.2 | 0.279 |

Similarly, the oxygen spin-orbit spectrum for all the samples can be fitted by two peaks (Figs. 4(a-d)). O 1s peak for lower BE value at 531.2 eV ($S_1$) and 531.3 eV ($S_2$, $S_3$, $S_4$) are attributed to O in lattice,[34] and that of higher BE values at 532.5 to 534 correspond to O from $SiO_2$[36,37] or absorbed O species. Significant increase in the amount of lattice O and $V^{5+}/V^{4+}$ ratio from sample $S_1$ to sample $S_4$ indicates that increasing gas partial pressure leads to increasing oxygen excess (or equivalently increasing $V_v$ concentrations, $n_V$). We have already discussed how V vacancy is consistent with the observed strain in the structural studies using GIXRD (Fig. 1) and hardening of phonon modes seen in Raman spectroscopy (Fig. 3).

The $V^{4+}$ $2p_{3/2}$ and O 1s peak are shifted towards higher BE in sample $S_2$, $S_3$, and $S_4$ with respect to sample $S_1$. Higher oxidation states are reported to have larger chemical shifts of spin-orbit levels than lower oxidation states with respect to neutral atom.[31] In sample $S_2$ to $S_4$, the existence of both $V^{4+}$ and $V^{5+}$ states is responsible for shift in binding energies towards higher energy values. In the $V^{5+}$–$V^{4+}$ chain, $V^{5+}$ is more electronegative due to the induced localized holes; therefore, V–O hybridization and strong V–V interaction result in the shifts of O 1s and $V^{4+}$ 2p core levels toward higher binding energies.[31] A plot of the $T_c$ and Raman active modes frequencies vs. $V^{5+}/V^{4+}$ ratios, which is considered proportional to $n_V$ (as discussed earlier) for all four samples are shown in Figure 5.

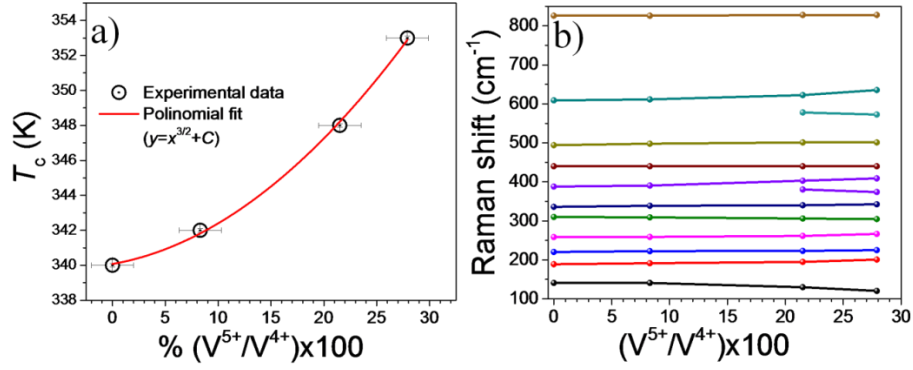

**Fig. 5.** (a) Variation of transition temperature with $V^{5+}/V^{4+}$ ratios ($\propto n_V$) with a polynomial fitting (red curve) exponent of 3/2 and a constant value $C$ corresponding to $T_C$ at $n_V$ =0 (340K). Bars denote errors in *x* and *y* axes. (b) Raman shift with $V^{5+}/V^{4+}$ ratios for the four samples.

The transition temperature increases with $V^{5+}/V^{4+}$ ratio of the samples $S_1$ to $S_4$ with a polynomial fitting exponent of 3/2 (Fig. 5(a)). Additionally, most of the Raman modes, except for the lowest frequency one (red shifted by an amount of 20 cm$^{-1}$), are blue shifted with increase in $V^{5+}/V^{4+}$ ratio (Fig. 5(b)), which confirms the role of $V_V$ concentration in increasing local strain.

*Role of finite sized Heisenberg spin ½ chain in modifying $T_C$*

Insulating and nonmagnetic nature of M1 phase of $VO_2$ suggests that it may be a typical Peierls insulator with all the V chains dimerized.[6] It is also found that uniaxial strain,[14] or doping of trivalent metals in $VO_2$,[5-7] leads to M2 phase, where V ions in alternate chains are only dimerized. Since M2 phase is also insulating, dimerization alone can not drive the MIT. Thus, whether electron-phonon coupling or strong electron-electron correlation is responsible for triggering the MIT accompanied by SPT in $VO_2$ is still under dispute. Zylbersztejn and Mott,[10] Sommers and Doniach,[38] and Rice *et al.*[26] suggested that Coulomb repulsion plays a major role in opening the energy gap in the insulating phases of $VO_2$. Qualitative electronic structures of $VO_2$ were proposed long ago by Goodenough,[11] where the *d* levels of the V ions splits into lower lying $t_{2g}$ states and higher energy $e_g$ states. The tetragonal crystal field further splits the $t_{2g}$ multiplet into an $a_{1g}$ state and a $e_g^\pi$ doublet (schematic in the supplementary information Fig. S2). The orbital nature of $a_{1g}$ states are such that the hopping integral between V ions within a chain along $c_R$ axis is larger compared to hopping in other two transverse directions (between V chains). In the monoclinic phases the dimerization and tilting of the V-V pairs leads to two effects. First, the $a_{1g}$ band is split into a lower-energy bonding combination and a higher-energy antibonding one. Secondly, the tilting of the pairs increases the overlap of these states with O states, pushing the $e_g^\pi$ states to higher energy. In Goodenough's picture, the single *d* electron occupies the $a_{1g}$ bonding combination (half-filled band). In the metallic R phase, all three ($a_{1g}$ $a_{1g'}$ and $e_g^\pi$) bands are partially occupied, whereas in the insulating phase only $a_{1g}$ is half-filled and overlapping $a_{1g'}$, $e_g^\pi$ bands are empty with a gap between them of the order of 0.6 eV (supplementary Fig. S2). It is important to note that density functional theory (DFT) with local density approximation (LDA)[12] does not find any gap between $a_{1g}$ and overlapped $a_{1g'}$, $e_g^\pi$ bands, even including structural distortion in the calculation, as imagined by Goodenough.[11] This shows that SPT may not be the driving force behind the MIT. On the other hand, LDA along with cluster dynamical mean field theory (DMFT)[39] does manage to open up a gap (though much smaller than experimental value), which increases with local Coulomb correlation strength. This shows that Coulomb correlation may be sufficient to cause level repulsion (separation) between $a_{1g}$ and overlapped $a_{1g'}$, $e_g^\pi$ states, leading to an orbital selection and a half-filled $a_{1g}$ band. The orbitals of the half-filled $a_{1g}$ band are such that, they have appreciable hopping only along the $c_R$-axis. So the reduction from three-band to one-band coincide with a reduction of dimensionality from three to one. One-dimensionality of the $a_{1g}$ band also leads to increase in effective local screened Coulomb repulsion energy (Hubbard $U$), and drives the chains towards Mott-Hubbard insulating phase (non-interacting, insulating Heisenberg spin ½ chains). Again, because of one-dimensionality, Heisenberg spin ½ chains are susceptible to spin-Pierls (dimerized phase) transition accompanying SPT. The electronic transition goes successively from, three-band, 3-D metal to a single half-filled band, then to 1-D Heisenberg chain and finally to spin-Peierls insulator. Reduction to 1-D due to orbital selection is the reason for SPT and MIT to happen at the same time, in a discontinuous fashion.

In M1 phase, all V ions are paired into dimers that are tilted away from $c_R$-axis. Excess oxygen, or equivalently $V_V$ leads to sites with $V^{5+}$ or $d^0$ states (holes in the spin ½ chains). If the $d^0$ sites are pinned at the nearest neighbors of $V_v$ sites,[31] they effectively cuts off the 1-D spin chains into smaller fragments at nanoscale. The average size of the chains decrease with increase in $V_v$ concentration. Spin-Peierls transition temperature in finite size system is larger than the transition temperature in the infinite system. The correlation length of order parameter near phase transition temperature scales as, $\xi(T) \propto (T-T_{c0})^{-\nu}$ where $\xi(T)$ is the correlation length at temperature $T$, and $T_{c0}$ is the thermodynamic (infinite system limit) transition temperature and $\nu$ is the critical exponent. In a thermodynamic transition, correlation length diverges at transition temperature. In finite size system, when the correlation length becomes of the order of the system size '$L$' at a temperature $T_c > T_{c0}$, the phase transition occurs. Substituting $T=T_c$ and $\xi(T)= L$ (where $L$ being the average length of the Heisenberg chains) in the scaling equation, we get $(T_c - T_{c0}) \propto L^{-1/\nu}$. Since average size of the chains is inversely proportional to the $V_V$ concentrations we get $(T_c - T_{c0}) \propto n_V^{1/\nu}$. It may be noted that $n_V$ is proportional to the experimentally measured ratio of $V^{+5}$ and $V^{+4}$ ions (Table I). From the experimental fit we find $\nu=⅔$ (with $C=T_{c0}$=340K in Fig. 5(a)), which matches exactly with the theoretically predicted value of critical exponent for dimerization transition in finite size spin ½ Heisenberg chain.[42-44] Increase in the $T_c$ value with decrease in sample size was also reported by few earlier workers in small particles,[40] and in thin films,[41] invoking presence of strain in the system. In our case the particle size does not change, but the $V_v$ concentration changes with increase in oxygen exposure leading to the reduction of the effective lengths of the Heisenberg chains as described earlier. Different concentration of $V_v$ essentially means different average length of the 1-D chains introducing finite size effect on $T_c$. The concentration of $d$-electrons at V sites is the only crucial parameter determining the $T_c$. On the other hand, unrelaxed strains in the materials do not influence $T_c$. However, it profoundly affects the structural type of the insulating phase. Local strains prevent V-V dimers from twisting in one chain, and consequently de-pairing the dimers in adjacent chains, promoting M2 or T phase instead of M1 phase at intermediate temperatures. The $T_c$ is reported to increase in $Cr^{+3}$, $Al^{+3}$, $Ga^{+3}$ ions doped $VO_2$.[5-7] These systems falls in the same universality class as our hole rich materials, because every trivalent metal ion also produces an adjacent $V^{+5}$ sites, just like every $V_v$ induces adjacent $d^0$ sites. It is also significant that trivalent atom substitution stabilizes M2 and T phases,[5-7] as observed in our materials, due to local lattice strains in the doped materials. So, the increase in $T_c$ with increase in $V_V$ concentration as well as appearance of intermediate phases during phase transition can be understood as a finite size effect in 1-D half-filled band. $W^{+6}$, $Mo^{+6}$, $Ta^{+5}$, $Nb^{+5}$ doping, on the other hand reduce the $T_c$.[24] We speculate that substitution with metal ion having valency more than four leads to doubly occupied $d^2$ vanadium sites. The doubly occupied sites can easily delocalize in the lattice (upper Hubbard band state) and one has to go down in temperature to arrest them into an insulating state, and thereby reducing the transition temperature in these materials. Thus the reduction from 3-band to 1-band along with the dimensional reduction from 3-D to 1-D, can explain both increase (decrease) in the $T_c$ value of $VO_2$ by doping with metal having valency less (more) than four.

**Conclusions**

The transition temperature ($T_c$) of MIT is recorded to increase with the increase in oxygen content for samples grown with different oxygen exposure. The presence of $V_v$ (with excess oxygen content) is found to generate unrelaxed local strain promoting T phase at the expense of M1 phase in the samples. The unrelaxed local strain however is not the driving force for the observed MIT. It is argued that the MIT is driven by strong electronic correlation. Mott type MIT and spin–Peierls type structural transition, both happens at the same time because of reduced one-dimensionality achieved in the lowest half-filled band. From the spectroscopic analysis, we argue that the insulating M1 phase should be viewed as a collection of V ions along $c_R$-axis of infinitely long Heisenberg spin ½ chains which are non-interacting to each other. The $V_V$ creates $d^0$ sites ($V^{5+}$) at the nearest neighbors introducing finite size scaling effect, as the $d^0$ sites reduce effective length of the Heisenberg spin ½ chains.

The increase in the $T_c$ value with the concentration of $V_V$ scales with a parametric value of ⅔, which matches with the reported scaling parameter of dimerized finite sized Heisenberg spin ½ chains at nanoscale confirming our phenomenological model for the increase in the $T_c$ value with defect doping. In fact, both the increase and decrease of $T_c$ value by doping can also be explained from our reduced 1-D band model, for the first time. Moreover, the controversy between Peirls and Mott-Hubbard in driving the MIT of $VO_2$ phases can also be resolved using our phenomenological model of reduced 1-D Heisenberg spin ½ chains.

## Acknowledgments

We thank A. K. Prasad, Avinash Patsha, A. K. Sivadasan, and Santanu Parida of SND, IGCAR for their valuable suggestions.## Notes and references

**Table I:** Binding energy values of $V^{4+}$ and $V^{5+}$ and $V^{5+}/V^{4+}$ ratio for samples $S_1$ to $S_4$

| Sample | $V^{4+}$ (in eV) | $V^{5+}$ (in eV) | $V^{5+}/V^{4+}$ |
|--------|------------------|------------------|-----------------|
| $S_1$  | 516.2            | -                | 0               |
| $S_2$  | 516.3            | 518.4            | 0.083           |
| $S_3$  | 516.3            | 518.3            | 0.215           |
| $S_4$  | 516.3            | 518.2            | 0.279           |

## Figure captions

**Fig. 1.** GIXRD spectra of the pristine samples (a) $S_1$ and (b) $S_4$ indicating crystallographic (*hkl*) planes of the corresponding phases.

**Fig. 2.** (a) Raman spectra of the pristine samples, $S_1$ to $S_4$ with proper symmetry notations, (b) Raman modes correspond to V-V chain ($\omega_1$ and $\omega_2$) for all the four samples and (c) characteristic peak of $VO_2$ ($\omega_0$) corresponds to V-O stretching for the samples.

**Fig. 3.** Raman spectra of the samples (a) $S_1$, (b) $S_2$ (c) $S_3$ and (d) $S_4$ as a function of temperature. Underlined temperatures denote the corresponding transition temperatures and phase reversibility. Peak observed at 521 cm$^{-1}$ in (b) for sample $S_2$ above 340K correspond to Si substrate. Solid and dashed arrows denote the increase and decrease in temperature, respectively.

**Fig. 4.** High resolution XPS spectra of sample (a) $S_1$, (b) $S_2$, (c) $S_3$, and (d) $S_4$ with corresponding electronic transition of different elements.

**Fig. 5.** (a) Variation of transition temperature with $V^{5+}/V^{4+}$ ratios ($\propto n_V$) with a polynomial fitting exponent of 3/2 and a constant value C corresponding to $T_C$ at $n_V$ =0 (340K). Bars denote errors in *x* and *y* axes. (b) Raman shift with $V^{5+}/V^{4+}$ ratios for the four samples.

TOC

# Role of 1-D finite size Heisenberg chain in increasing metal to insulator transition temperature in hole rich VO$_2$

Raktima Basu, Manas Sardar, Santanu Bera, P. Magudapathy, Sandip Dhara

The observation of metal to insulator transition and structural phase transition at the same time in VO$_2$ can be explained from our phenomenological model of reduced 1-D Heisenberg spin ½ chains along with the reported increase (decrease) in $T_c$ value of VO$_2$ by doping with metal having valency less (more) than four, for the first time.

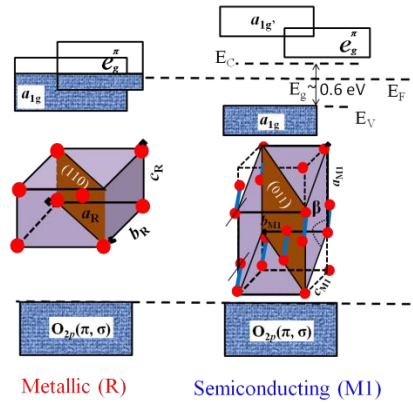

# Supplementary Information

# Role of 1-D finite size Heisenberg chain in increasing metal to insulator transition temperature in hole rich VO$_2$

Raktima Basu,[1,*] Manas Sardar,[2] Santanu Bera,[3] P. Magudapathy,[2] Sandip Dhara[1,*]

[1]Nanomaterials and Sensor Section, Surface and Nanoscience Division, Indira Gandhi Centre for Atomic Research, Homi Bhabha National Institute, Kalpakkam-603102, India

[2]Materials Physics Division, Indira Gandhi Centre for Atomic Research, Kalpakkam-603102, India

[3]Water and Steam Chemistry Division, BARC Facility, Kalpakkam-603102 & Homi Bhabha National Institute, Mumbai-400094, India

E-mail: raktimabasu14@gmail.com ; dhara@igcar.gov.in

Supplementary Figures

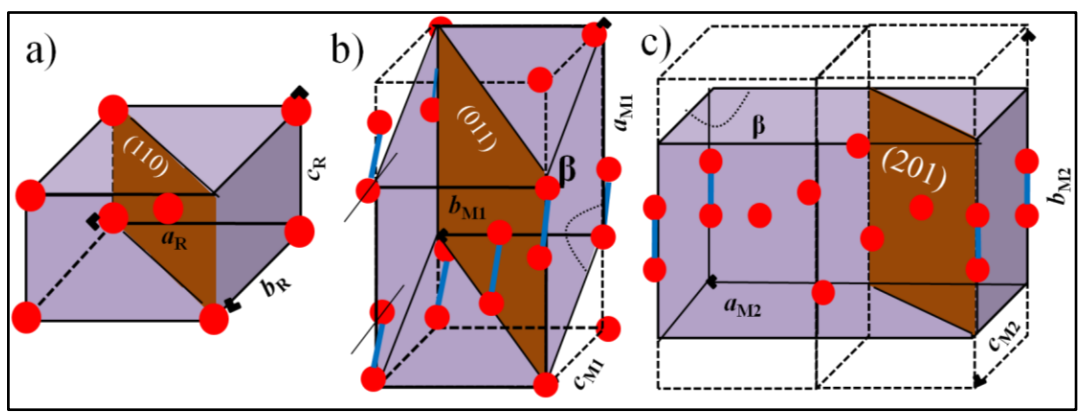

**Fig. S1.** Schematic structural diagram of different phases of VO$_2$ a) rutile tetragonal, R; b) monoclinic, M1; and c) monoclinic, M2. Red balls denote vanadium atoms, each surrounded by oxygen octrahedra with principle axis perpendicular to the shaded plane. Oxygen atoms are not shown in the figure.

Figure S1 shows the schematic diagram of the different structural phases of VO$_2$. In each phase of VO$_2$ there are two interpenetrating parallel chains of V surrounded by six O atoms forming distorted octahedron. The oxygen atoms are not shown in the figure. The high temperature rutile tetragonal (R) phase (Fig. S1a) is the most symmetric one with lattice parameter $a_R = b_R = 4.555$ Å, $c_R = 2.853$ Å, $\alpha = \beta = \gamma = 90°$.[1] In the monoclinic M1 phase (Fig. S1b), the V atoms form pair (dimerised) and the pairs tilt along the $c_R$ axis making the unit cell double of the unit cell in R phase with the approximate crystallographic relationship as $a_{M1} \leftrightarrow 2c_R$, $b_{M1} \leftrightarrow a_R$, and $c_{M1} \leftrightarrow b_R - c_R$.[2,3] In monoclinic M2 phase (Fig. S1c), the one set of V chains along the $c_R$ axis pair without twisting, while the V ions in the nearest neighbor V chains, do not pair but twist away from $c_R$ axis. The crystallographic relationship with the R phase is $b_{M2} \leftrightarrow 2c_R$, $a_{M2} \leftrightarrow 2a_R$, and $c_{M2} \leftrightarrow -b_R$.[3] The triclinic T phase is intermediate between the M1 and M2 phases, in which paired V chains in the M2 phase are reported to be twisted slightly.[4]

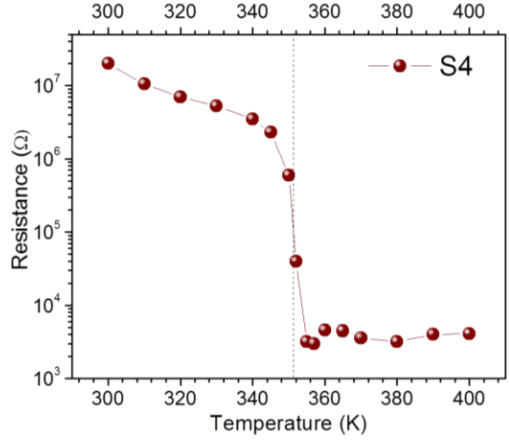

**Fig. S2.** Typical electrical transport measurement for the VO$_2$ sample (S4) showing a drop in resistance of four orders indicating metal insulator transition ~ 355K (vertical dashed line). Symbol is for the experimental data and connecting line is a guide to eyes.

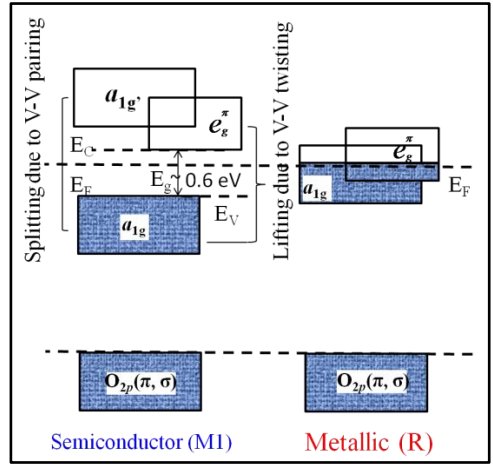

**Fig. S3.** Schematic electronic band structure for semiconducting and metallic phase of VO$_2$

The electronic band structure of VO$_2$ in the metallic rutile phase and insulating monoclinic M1 phase are shown in Figure S5. The $d$ levels of the V ions splits into lower lying $t_{2g}$ states and higher energy $e_g$ states. The tetragonal crystal field further splits the $t_{2g}$ multiplet into an $a_{1g}$ state and an $e_g$ doublet. In the insulating phase, pairing of vanadium atoms in the $c_R$ direction splits the $a_{1g}$ bands into lower (bonding, $a_{1g}$) and upper (antibonding, $a_{1g'}$) bands. Furthermore, the twisting of V-V pairs enhances the V$d$-O$p$ hybridization and thereby raises $e_g^\pi$ band above the Fermi level, which in turn opens up a gap of ~ 0.6 eV.[3]

**Notes and references**